\documentclass[twocolumn,superscriptaddress]{revtex4-1}
\usepackage{graphicx,float}
\usepackage{latexsym}
\usepackage{epstopdf}
\usepackage{amssymb,amsmath}
\usepackage{color}
\usepackage{multirow}
\usepackage{hyperref}
\usepackage{kantlipsum}
\makeatletter
\let\Hy@linktoc\Hy@linktoc@none
\makeatother

\hypersetup{colorlinks=true,linkcolor=blue}
\usepackage{verbatim}
\usepackage{bm}

\newcommand*\Laplace{\mathop{}\!\mathbin\bigtriangleup}
\begin{document}

\title{Anomalous phase diagram of the elastic interface with non-local hydrodynamic interactions in the presence of quench disorder}
\author{Mohsen Ghasemi Nezhadhaghighi} 
\email{mgnhaqiqi@gmail.com} 
\affiliation{Department of Physics, School of Science, Shiraz University,
Shiraz 71946-84795, Iran} \date{\today}

\begin{abstract}

We investigate the influence of quenched disorder on the steady states of driven systems of the elastic interface with non-local hydrodynamic interactions.
The generalized elastic model (GEM), which has been used to characterize numerous physical systems such as polymers, membranes, single-file systems, rough interfaces, and fluctuating surfaces, is a standard approach to studying the dynamics of elastic interfaces with non-local hydrodynamic interactions. The criticality and phase transition of the quenched generalized elastic model (qGEM) are investigated numerically, and the results are presented in a phase diagram spanned by two tuning parameters. We demonstrate that in 1-d disordered driven GEM, three qualitatively different behavior regimes are possible with a proper specification of the order parameter (mean velocity) for this system. In the vanishing order parameter regime, the
steady-state order parameter approaches zero in the thermodynamic limit. A system with a non-zero mean velocity can be in either the continuous regime, which is characterized by a second-order phase transition, or the discontinuous regime, which is characterized by a first-order phase transition. The focus of this research was to investigate at the critical scaling features near the pinning-depinnig threshold. The behavior of the quenched generalized elastic model at the critical “depinning” force is explored.
Near the depinning threshold, the critical exponent obtained numerically.

\end{abstract}

\maketitle

\section{Introduction}
The study of universal scaling behaviors associated with the nonequilibrium critical phenomena is an attractive and fascinating field of statistical physics that has attracted a considerable attention in recent years  \cite{kardar1998nonequilibrium,fisher1998collective,amaral1995scaling,
kawamura2012statistical,gros2014}. 
 Indeed, it is expected that a wide range of models at the critical point could well be characterized by the same universal parameters, as is well known from equilibrium critical phenomena \cite{odor2004universality}. 
Is it possible to derive these parameters to determine the universality of critical phase transitions in out-of-equilibrium models? Recent studies have focused on the dynamical characteristics of a vast range of problems, including fracture propagation in solids \cite{gao1989first,schmittbuhl1995interfacial,tanguy1998individual,alava2006statistical,priol}, charge density waves in anisotropic
conductors \cite{fisher1985sliding,gruner1988dynamics}, vortices in type-II superconductors \cite{blatter1994vortices}, domain walls
in ferromagnetic \cite{lemerle1998domain} or ferroelectric \cite{tybell2002domain} systems, 
the contact line of a fluid drop on a disordered substrate \cite{cieplak1988dynamical,moulinet2004width,alava2004imbibition,prim2019,rabbani},
the deformation of crystals \cite{alava2006statistical}, crackling noise in a wide range of physical systems from magnetic materials to paper crumpling
 \cite{sethna2001crackling,bonamy2008crackling}, friction and lubrication \cite{cule1996tribology,vanossi2011modeling}, 
the motion of geological faults \cite{fisher1997statistics}, tumour growth \cite{bru2004pinning,moglia2016}, and many others. This diverse set of processes may be described as an extended elastic manifold driven over quenched disorder, which has a complicated dynamics that includes non-equilibrium phase transitions.

The competition between the deformation induced by quenched disorder (induced by the presence of impurities in the host environment) and the elastic material's response to an applied driving force is the key factor determining their dynamical behaviour in all of these complex non-linear systems.

The "depinning transition" phenomenon is a significant result of this competition \cite{surface2}. 
In the absence of an external driving force $F$, 
the system is disordered but it does not move and remains \textit{pinned}
by the quench disorder. When the external force is increased from zero, 
the elastic object \textit{unpins} and reaches a finite steady-state velocity \cite{fisher1998collective}.
This describes the critical phase transition of the elastic interface 
at the critical force $F = F_c$, where the driving force $F$ plays the
role of the control parameter and the mean velocity $v$ is the
order parameter \cite{kardar1998nonequilibrium}. Note that the critical value of the external force $F_c$ is not
universal and its value depends on the details of the model. 
The steady-state average velocity follows a power-law characteristic as $v\sim (F-F_c)^\theta$ while approaching the critical point from above, where $\theta$ is a universal parameter. Other measures, such as the local width, the correlation functions, the correlation length, and the structure factor, may be used to extract the exponents associated to the criticality of the elastic interface. These techniques have been extensively used to investigate the self-affine surface structure's scaling properties \cite{surface2,mckane2013scale,krug1997origins}.

Consider a single-valued function $u(\textbf{x},t)$ that describes an elastic interface.  
The global surface width $W= \sqrt{\langle (u(\textbf{x},t) - \langle u \rangle_{\textbf{x}})^2 \rangle_{\textbf{x}}}$, is the simplest quantity used to characterize the scaling characteristics of elastic interfaces near the critical point,  which is defined as the standard deviation around the mean position. 
For a finite system of size $L$, the roughening of $u$ from a 
flat initial condition, scales as:  
\begin{eqnarray}
W(L,t) \sim t^\beta f(L/t^{1/\nu}),
\end{eqnarray}
where the exponents $\beta$  and $\nu$ are called the growth and the dynamical exponent. The scaling 
function $f(x)$ is such that $f(x)\sim \textrm{const}$ for $x\gg 1$, and $f(x) \sim x^{\zeta_{g}}$ 
for $x\ll 1$ (the exponent $\zeta_g$ is known as the global roughness exponent). Finite size effects, as expected, occur when $t^{\times}_W \sim L^\nu$. The self-affine
scaling relates now $\zeta_g$, $\beta$, and the dynamical exponent $\nu$ through $\nu = \zeta_g/\beta$ \cite{surface2}.

The average velocity, which corresponds to the order parameter of the pinning-depinning transition of a driven interface, may be assumed to be a homogeneous function of time $t$ and $|F-F_c|$, similar to critical phenomena as:
\begin{eqnarray}
v(t,F) \sim t^{-\sigma} g(|F-F_c|t^{\sigma/\theta}),
\end{eqnarray}
where $\sigma$ is a universal scaling exponent. For $F > F_c$ there is a crossover time-scale, $t^{\times}_v\sim |F-F_c|^{\theta/\sigma}$ between two regimes: $g(x) \rightarrow \textrm{const}$ for $t\ll t^{\times}_v$ and $g(x)\sim x^\theta$ for $t \gg t^{\times}_v$.  

The equilibrium configuration of an elastic rough interface in the critical point is expected to be self-affine, and the tow-point correlation function is supposed to obey the scaling form
\begin{eqnarray}
\label{coorelation function}
C(r)=\langle [ u(\textbf{x})- u(\textbf{x}^\prime) ] ^2\rangle \sim |\textbf{x}-\textbf{x}^\prime |^{2\zeta_l},
\end{eqnarray}
where $\zeta_l$ is the local roughness exponent. 

Various experimental, analytical, and numerical works have been proposed to compute the critical exponents $\theta$, $\beta$, $\zeta_g$, $\zeta_l$, and $\sigma$ characterizing the ``pinning-depinning'' phase transition, in a similar fashion to the equilibrium critical phenomena.

The purpose of this research is to describe and investigate the statics and dynamics of a generalized model for the investigation of a range of other reported phenomena in which the pinning-deppinig phase transition may occur.
The paper is organized as follows. Section \ref{Definition sec} introduces the model. 
Section
\ref{Numerical method sec} describes the numerical formalism. 
In Sec. \ref{Numerical results sec}  we discuss our findings. In the final section, we summarize the obtained results and our conclusions.

\section{Definition of the model}\label{Definition sec}
Despite the significant variations in theoretical models, many of the computations were performed using the linear assumption of the elasticity $u(\textbf{x},t)$. The following equation can be used to explain the motion of an interface in an isotropic disordered material at this level of precision \cite{fisher1998collective}:
\begin{eqnarray}
\label{fisher_formula}
\frac{\partial u(\textbf{x},t)}{\partial t} = F  + f_p(\mathbf{x},u(\mathbf{x},t)) -\mathcal{K}\left[ u(\mathbf{x},t) \right]~,
\end{eqnarray} 
where $F$ is a uniform external force which is also the control parameter and $f_p$ representing
the ``non-thermal'' quenched random forces due to the randomness and impurities of the heterogeneous medium. 
The quenched random noise $f_p(\mathbf{x},u(\mathbf{x},t))$, can be taken to have zero mean satisfying the relation 
$\langle f_p(\mathbf{x},u) f_p(\mathbf{x}^\prime,u^\prime) \rangle = 2D \delta (\mathbf{x}-\mathbf{x}^\prime) \mathcal{R} (u-u^\prime)$, where $\mathcal{R}(u-u^\prime)$ assumed to decay rapidly for large values of its argument. The final term $\mathcal{K}\left[ u(\mathbf{x},t) \right]$ in Eq. (\ref{fisher_formula}) describes the elastic forces between different parts. It has the form
\begin{eqnarray}
\mathcal{K}\left[ u(\mathbf{x},t) \right] = \int d^D\textbf{x}^\prime \int dt^\prime \mathcal{J}(\textbf{x}-\textbf{x}^\prime,t-t^\prime)\nonumber \\
\times \left[ u(\textbf{x}^\prime,t^\prime) - u(\textbf{x},t) \right],
\end{eqnarray}
where $D$ is the space dimension and $\mathcal{J}(\textbf{x}-\textbf{x}^\prime,t-t^\prime)$  is defined as the propagation kernel to transmit the stress on the interface from its elasticity. Moreover, systems
with short range elasticity of the interface are
characterized by $\mathcal{J}(\textbf{x},t)\propto \delta (t) \nabla^2 \delta (\textbf{x})$ \cite{fisher1998collective}.

Theoretical studies on quenched disordered systems, such as a contact line of a liquid meniscus on a disordered substrate \cite{rosso2002roughness,moulinet2004width}, 
crack propagation \cite{rosso2002roughness,laurson2010avalanches} 
and solid friction \cite{moretti2004depinning}, have shown that it is possible to express
the kernel $\mathcal{K}[u]$ in a long-range form 
\begin{eqnarray}
\label{long range kernel}
\mathcal{K}\left[ u(\mathbf{x},t) \right]\propto \int d^D\textbf{x}^\prime \frac{u(\textbf{x},t)-u(\textbf{x}^\prime,t)}{|\textbf{x}-\textbf{x}^\prime|^{D+z}},
\end{eqnarray}
where the exponent $z$ is a variable that depends on the model chosen to represent the elastic interface \cite{tanguy1998individual}. The most important aspect of the singular integration Eq. (\ref{long range kernel}) is that it may be used to rewrite the elastic force $\mathcal{K}[u]$ as 
\begin{eqnarray}\label{fractional fisher}
\mathcal{K}\left[ u(\mathbf{x},t) \right] = \left( - \Laplace \right) ^{z/2} 
u(\mathbf{x},t),
\end{eqnarray}
where $\left( - \Laplace \right) ^{z/2}$ is the fractional Laplacian 
defined by its Fourier transform 
$\widehat{\left( - \Laplace \right)} ^{z/2} \Phi (\textbf{k}) = 
|\textbf{k}|^z \widehat{\Phi }(\textbf{k})$ \cite{fractiobaloperators}.
According to Eqs. (\ref{fisher_formula}) and (\ref{fractional fisher}),
one can rewrite the Eq. (\ref{fisher_formula}) as follows:
\begin{eqnarray}
\label{generalized_fisher_formula}
\frac{\partial u(\mathbf{x},t)}{\partial t} =F +f_p(\mathbf{x},u(\mathbf{x},t))-
\left( - \Laplace \right) ^{z/2} u(\mathbf{x},t) 
~.
\end{eqnarray}

It is indeed worth mentioning that the dynamics given by Eq. (\ref{generalized_fisher_formula} are essentially generalizations of the quenched Edwards-Wilkinson (qEW) and quenched Mullins-Herring (qMH) equations, which are the simplest and most often used equations to explain the interface pinning-depinning transition in quenched random media, with $z=2 \textrm{ and } 4$, respectively.

Many researches have been carried on the qEW and qMH equations, as well as the related models. Early studies investigated numerically the crucial characteristics of the qEW equation \cite{leschhorn1993interface},
and it has been the subject of many
theoretical and numerical studies in recent years 
\cite{ramasco2000generic,rosso2001origin,lacombe2001force,rosso2003depinning,kolton2006short,kolton2009universal}.
Recently, a novel and very efficient approach investigated the qEW equation's depinning threshold and critical exponents \cite{ferrero2013numerical,ferrero2013nonsteady}.
The scaling properties of the qMH equation at the critical point of the pinning-depinning transition have been quantitatively explored \cite{lee2000growth,lee2006depinning,boltz2014depinning}.
It is worth noting that, for the 
so-called space-fractional quenched equation Eq. (\ref{generalized_fisher_formula}), the scaling hypothesis has been  established in Ref. \cite{xia2012depinning} 
(The fractional power $z$ is expected to be in the range $1.5 \leqslant z\leqslant 2$). The Grunwald-Letnikov form of a fractional derivative has been used to discretize the space-fractional quenched equation, which is essentially an integro-differential equation, as noted in Ref.\cite{xia2012depinning}.

Despite the success of
Eqs. (\ref{fisher_formula}) and (\ref{generalized_fisher_formula}) 
in describing the dynamics of elastic interfaces driven through a disordered medium, this toy model had one weakness: hydrodynamic interactions were not included. This is the case, for instance, of polymers \cite{haidara2008competitiv,d2010single},
membranes \cite{nissen2001interface,verma2014rough}, the dynamics of colloid suspensions, macromolecular solutions and multicomponent systems \cite{clague1996hindered,miguel2003deblocking,cui2004anomalous,Sbragaglia2014,stannard2011dewetting}. 
Because of the long-range hydrodynamic interaction, the dynamical behavior of these systems is correlated via flows.

The generalized elastic model (GEM), proposed in Ref. \cite{Taloniprl}, is a suitable linear model that may capture the essence of criticality and phase transition (see \cite{Talonipre,Taloniepl,Taloniperturb,Talonirev} for more details). In this case, we used this model in the presence of a quenched disorder. The quenched form of the generalized elastic model (qGEM) is represented by the stochastic linear integrodifferential equation shown below
\begin{eqnarray}
\label{qGEM}
\frac{\partial u(\mathbf{x},t)}{\partial t} =F +
\int d^d x^\prime \Lambda(\vert \mathbf{x}-\mathbf{x}^\prime\vert)\frac{\partial^z}
{\partial |\mathbf{x}^\prime | ^z}u(\mathbf{x}^\prime,t)\nonumber \\
+f_p(\mathbf{x},u(\mathbf{x},t))
~,
\end{eqnarray}
where the dynamical variables of the system $u(\mathbf{x},t)$ describes 
an elastic interface driven through a disordered media. 
$F$ is the driving force on the interface and
$f_p$ represents the quenched pinning forces
which its distribution can be chosen Gaussian with the
first two moments, $\langle f_p (\mathbf{x},u)\rangle = 0$ and
$\langle f_p(\mathbf{x},u) f_p(\mathbf{x}^\prime,u^\prime)\rangle \propto \delta (\mathbf{x}-\mathbf{x}^\prime)\delta(u-u^\prime) $.
The hydrodynamic interaction term $\Lambda(\vert \mathbf{x}-
\mathbf{x}^\prime\vert)$, corresponds to the non-local coupling of 
different sites $\mathbf{x}$ and $\mathbf{x}'$.
Here, $\partial^z/\partial|\mathbf{x}|^z$ is the multidimensional
Riesz-Feller fractional derivative operator, which is  
defined via its
Fourier transform 
$\mathcal{F}\left\{\frac{\partial^z}{\partial|\mathbf{x}|^z}\Phi (\textbf{x})\right\}
\equiv-|\mathbf{k}|^z  \Phi (\textbf{k})$, 
immediately implies that the Riesz-Feller fractional derivative has 
the same meaning as the 
fractional Laplacian operator $\partial^z/\partial|\mathbf{x}|^z:=-\left( - 
\Laplace \right) ^{z/2}$ \cite{fractiobaloperators}.

\begin{figure}[t]
\begin{center}
\includegraphics[width=7cm,clip]{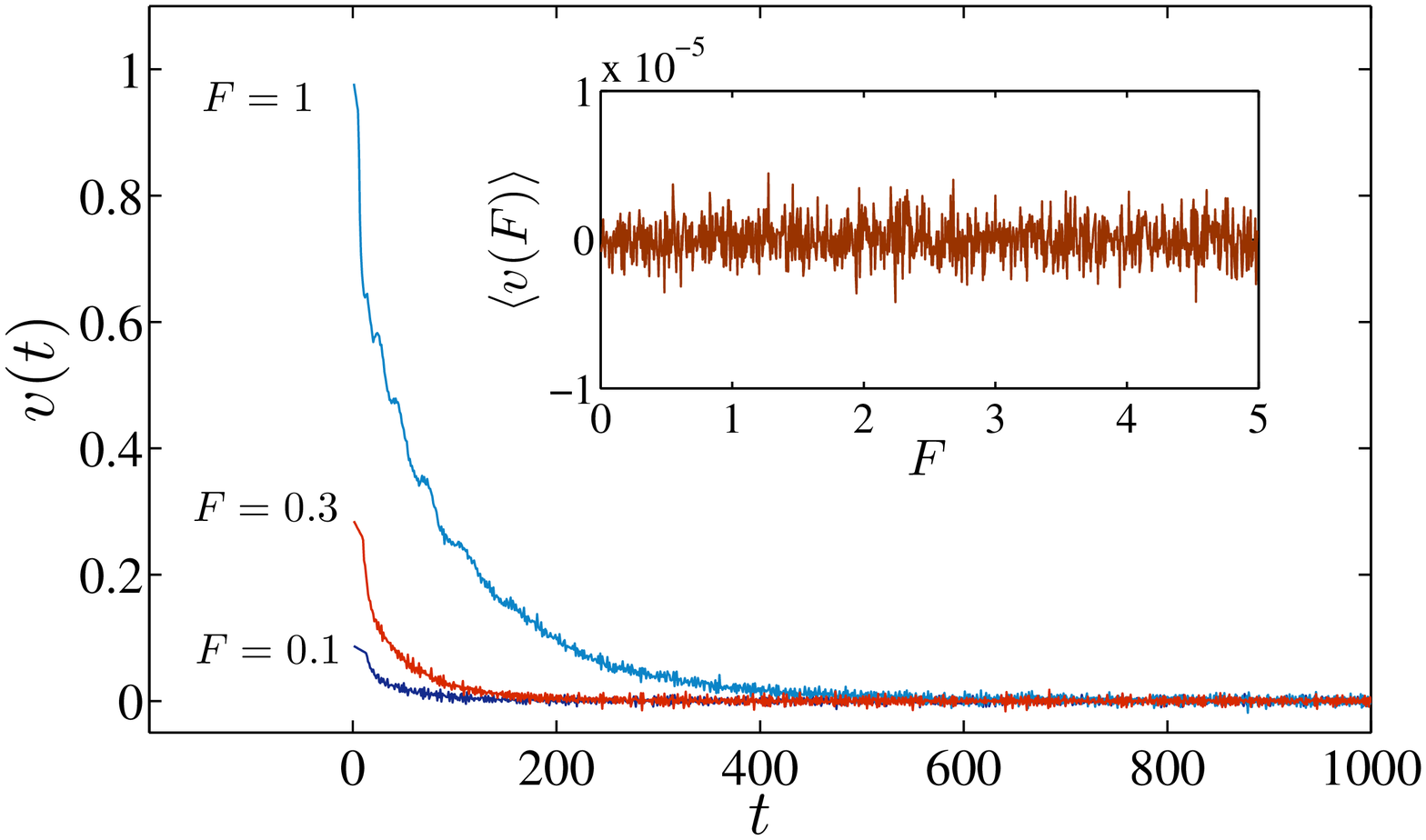}
\includegraphics[width=7cm,clip]{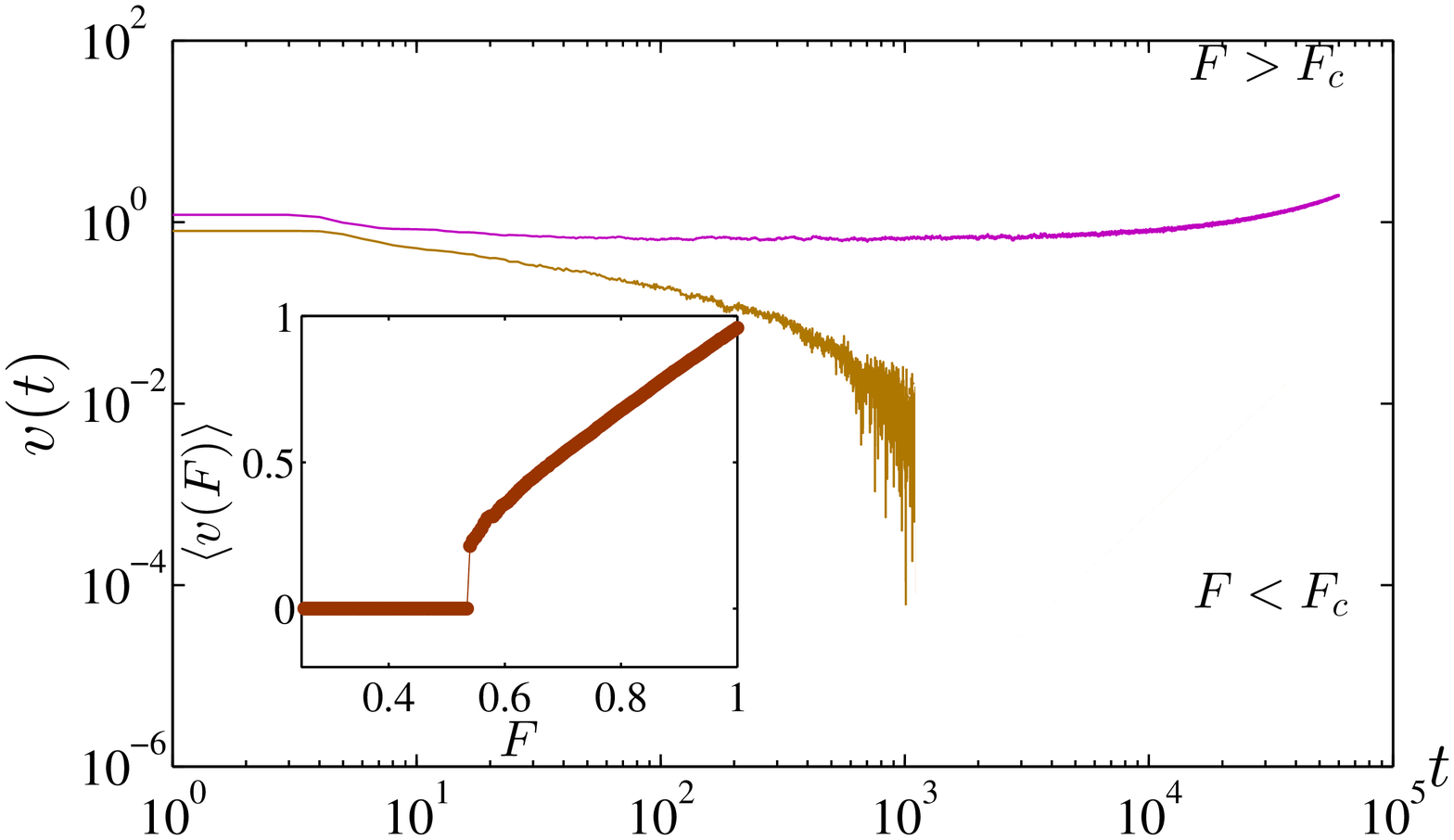}
\includegraphics[width=7cm,clip]{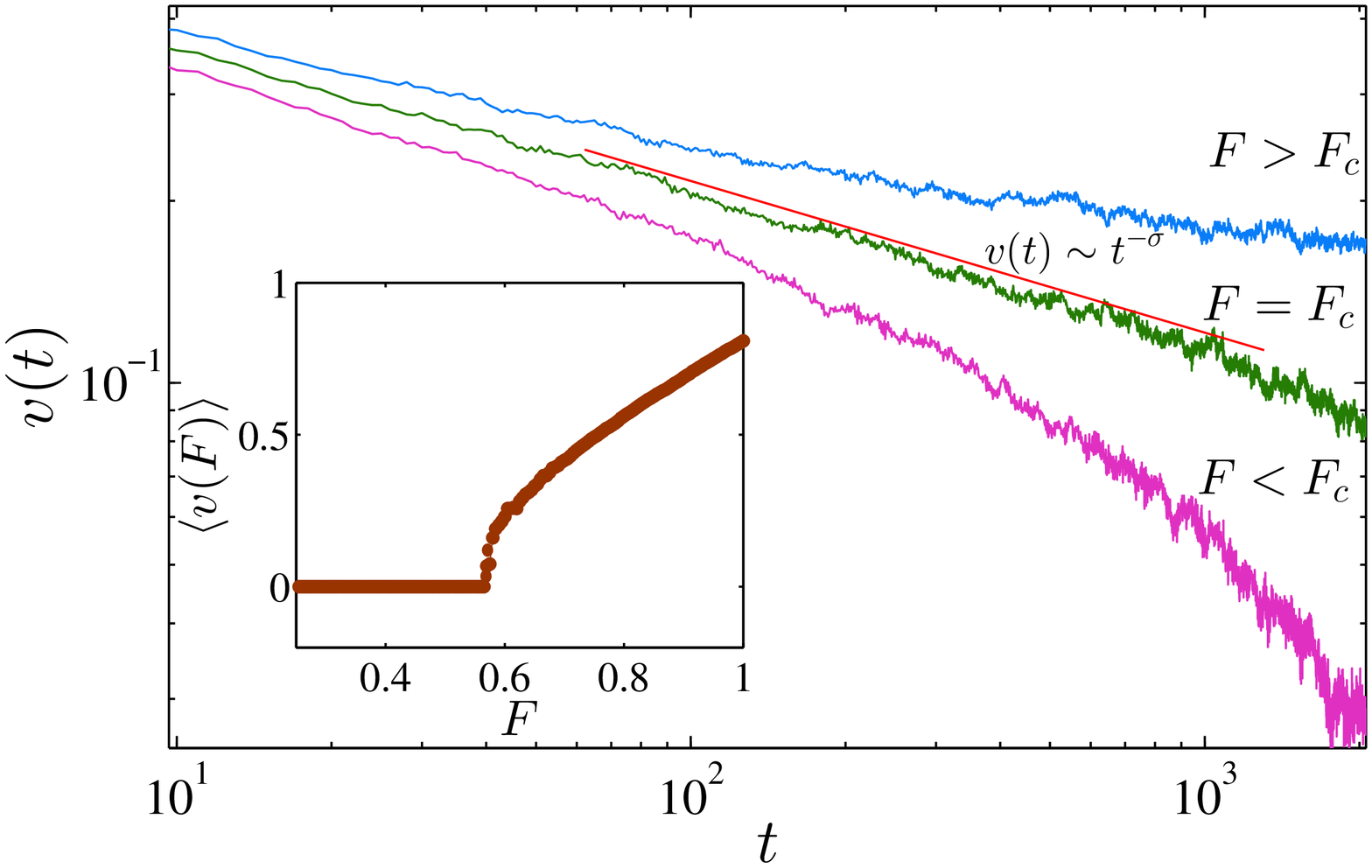}
\end{center}
\caption{(Color online)The numerical evaluation for the average velocity $v(t,F) = \frac{d}{dt} \langle \int  u(x,t)dx \rangle$ for the generalized elastic model with quenched disorder which corresponds to the order parameter of the pinning-depinning transition of the visco-elastic interface driven through a disordered media. The behavior of the order parameter is strongly influenced by the hydrodynamic interaction parameter $\alpha$ and the fractional power $z$. Top: The order parameter $v(t)$ as a function of time for three different values of the external force $F$. It goes to zero, when $t\to \infty$. The saturation values of the order parameter $v(F)$ is shown in the inset. Note that there is no phase transition for the values $\alpha = 0.5$ and $z=1.0$ in Eq. (\ref{qGEM}). Middle: The order parameter as a function of time and force for the so called $1$st order phase transition for the values $\alpha = 0.5$ and $z=3.0$ Bottom: The same analysis for  the values $\alpha = 0.5$ and $z=4.0$  shows an ordinary pinning-depinnig phase transition. Solid red line corresponds to the scaling relation $v(t) \sim t^{-\sigma}$ for the critical point $F=F_c$. 
}
\label{Figure:1}
\end{figure}

At this point a specification
of the hydrodynamic interaction kernel $\Lambda (\vec{r})$ is called for. 
For no fluid-mediated interactions, one may suppose that the friction kernel is local, $\Lambda (\vec{r})=\delta (|\vec{r}|)$ (examples). 
For systems having non-local interactions, such as membranes, polymers, or viscoelastic surfaces, where the hydrodynamic interactions take on a long-range power-law form, a different scenario 
\begin{eqnarray}\label{non local hydrodynamics kernel}
\Lambda (\vec{r}) \sim \frac{1}{|\vec{r}|^{\alpha}},
\end{eqnarray}
occurs (where $\frac{D-1}{2}<\alpha<D$) \cite{Taloniprl}. 
It should be emphasized that the $D$-dimensional Fourier transform of the 
hydrodynamic friction kernel Eq. (\ref{non local hydrodynamics kernel}) 
is given by $\Lambda (\textbf{q}) = A |\textbf{q}|^{\alpha -D}$ 
($A=\textrm{const}$). It's clear that the local hydrodynamic interaction corresponds to the act of taking $\alpha = D$ and $A=1$  
($\Lambda (\textbf{q}) = 1$) \cite{Talonipre}. 

The next section will present a detailed description of the discretization approach used to numerically explore the generalized elastic model in the presence of quenched disorder Eq. (\ref{qGEM}) for various values of the fractional order $z$ and the non-local hydrodynamic interaction strength $\alpha$.

\section{Numerical algorithm}\label{Numerical method sec}
We consider here the qGEM (\ref{qGEM}) in one spatial dimension $D=1$.  
The interface position $u(x_i,t_n)$ is specified on a lattice of size $L$, where $x_i= i\Delta x$ and $t_n = n\Delta t$ are defined with $i=0,\dots, L$ and $u_i^n$ is kept as a continuous variable.

To solve Eq (\ref{qGEM}) in discretized time and space, use the finite difference approximation to estimate the time derivative (forward Euler method): 
\begin{eqnarray}\label{time finite difference}
\frac{\partial u(x_i,t_n)}{\partial t} = \frac{u(x_i,t_{n+1})-u(x_i,t_n)}{\Delta t}~.
\end{eqnarray}
The discrete space Riesz-Feller fractional operator $\partial^z /\partial|x|^z$ in the Eq. (\ref{qGEM})
can be approximated using the matrix transform method proposed 
by Ili\'c \textit{et al} \cite{ilic2005numerical,ilic2006numerical}. 
Moreover, there are many other different numerical methods 
have been proposed to simulate such fractional operators \cite{Yang:thesis}.
Let us first consider the common notation for the Riesz-Feller derivative in terms of the 
Laplacian $\partial^z /\partial \vert {x} \vert^z:=-(-\Laplace)^{z/2}$ 
\cite{saichev1997fractional}. 
The matrix transform algorithm is based on the following definition: 
First consider the usual finite
difference scheme for Laplacian in one dimension
\begin{eqnarray}\label{fourier laplace}
\Laplace\phi(x) = \frac{1}{(\Delta x)^2} \lbrace \phi(x-\Delta x)-2\phi(x)+\phi(x+\Delta x) \rbrace,
\end{eqnarray} 
where $\lbrace \phi(x) \rbrace$ is the complete set of orthogonal functions. 
Using the Fourier transform 
$\phi(x) = \frac{1}{2\pi}\int \widehat{\phi} (q)e^{-iqx}dq$, 
the discretized Laplacian Eq. (\ref{fourier laplace}) in 
the Fourier representation can be rewritten as,
\begin{eqnarray}
\widehat{(\Laplace)}\phi(q) = -(2-2\cos (q\Delta x))\phi(q),
\end{eqnarray}
where $\Delta x$ corresponds to the lattice constant. 

One might start with the Fourier representation of the discretized Laplacian to approximate the Fourier representation of the discretized fractional Laplacian $(-\Laplace)$  as: $\lambda(q) = 2(1-\cos(q))$, and raise it to appropriate power, 
 $(2(1-\cos(q)))^{z/2}$. This technique has been invented by Ili\'c \textit{et al} 
 (for more details see Refs. \cite{ilic2005numerical,ilic2006numerical,Yang:thesis}).   

\begin{figure}[t]
\begin{center}
\includegraphics[width=8cm,clip]{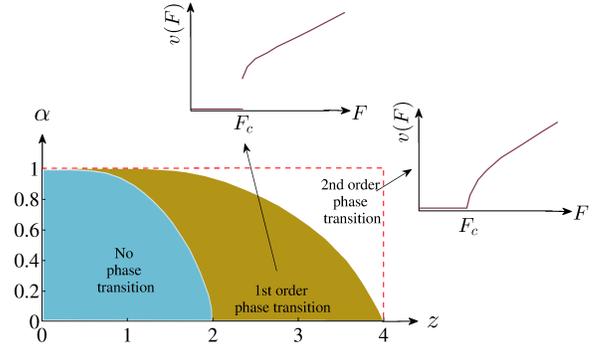}
\end{center}
\caption{(Color online)The phase diagram of the generalized elastic model with 
quenched disorder. There are three different regimes, depending on the values of
the parameters $z$ and $\alpha$ in Eq. (\ref{qGEM}):
The first regime is when $z\ll 4$ and $\alpha<1$ where there is no phase
transition between pinned and moving phases. The second regime is when  
$z < 4$ and $\alpha \gg 0$ where the order parameter of the system $v(F)$ 
as a function of the control parameter $F$ changes
continuously from zero to the non-zero values ($2nd$ order phase transition).
In the third regime the mean velocity $v(F)$ 
as a function of $F$ changes
discontinuously from zero to the non-zero values ($1st$ order phase transition).    
}
\label{Figure:2}
\end{figure} 
 
The matrix transform approach proposes that one can obtain the elements of the matrix representation of the Laplacian $\mathbb{A}_{l,m}=-\int_0 ^{2\pi}\frac{dq}{2\pi} 
(2-2\cos (qa)) e^{iq(l-m)}$ where $\mathbb{A}\equiv\text{tridiag}(1,-2,1)$.
The elements of the matrix $\mathbb{K}$, representing the discretized fractional Laplacian $-(-\Laplace)^{z/2}$ are then 
\begin{eqnarray} \label{HOLR} 
\mathbb{K}_{l,m}&= -\int_0^{2\pi} \frac{dq}{2\pi} e^{iq(l-m)} \left[ 2(1-\cos(q))\right]^{\frac{z}{2}}\nonumber \\
&=\frac{\Gamma(-\frac{z}{2}+n)\Gamma(z +1)}{\pi \Gamma(1+\frac{z}{2}+n)} \sin(\frac{z}{2}\pi),
\end{eqnarray}
where $n=\vert l-m\vert$, and fractional order $z\geq 1$. In the special case $z=2$ the $\mathbb{K}$ matrix is equal to the matrix $\mathbb{A}$ of a simple Laplacian. On the other hand, if $\alpha/2$ is an integer, then 
$\mathbb{K}(n) = (-1)^{\alpha-n+1}C_{\alpha,\frac{\alpha}{2}+n}$ for $n\leq\alpha/2$ and $\mathbb{K}(n)=0$ for $n>\alpha/2$, where $C_{\alpha,\frac{\alpha}{2}+n}$ are binomial coefficients \cite{zoia2007fractional}. 
 
Bringing together Eqs. (\ref{time finite difference}) and (\ref{HOLR})
and substitution into the Eq. (\ref{qGEM}) leads to the discrete version of 
the qGEM.
We employ the finite difference method to investigate the numerical 
discretization of Eq.(\ref{qGEM}), in the form
\begin{eqnarray}\label{discretqGEM}
u_i^{n+1} =u_i^n &+ \Delta t \lbrace   \frac{1}{(\Delta x)^z} \sum_{j= 0}^{L}\sum_{k= 0}^{L} \Lambda (\vert i-j\vert)\mathbb{K}_{j,k}u_k^n \nonumber \\
& +F +  f_p(x_i,u_i^n)\rbrace
 ~, 
\end{eqnarray}
where $u_i^n$ approximates the interface profile $u(x_i , t_n )$ at the $i$th lattice point and the $n$th time step. The lattice constant $\Delta x$ has been set equal 
to one and the grid steps  $\Delta t$ in time was chosen small enough to 
avoid numerical instabilities. 

In order to numerically generate quenched random field 
 $f_p(x_i,u_i^n)$, without loss of generality we assumed the continuous stochastic variables $u(x_i , t_n )$ are discretized into a finite numbers of integer values $[u_i^n/\epsilon]$ where $\epsilon \ll 1$ is an arbitrary small parameter and $[\dots]$ represents the bracket notation for the integer part of a given continuous variable.
Then the quenched random field $f_p$ is defined on a square array where each cell $[i,h]$ ($1\leqslant i \leqslant L$ and $h =[u_i^n/\epsilon ]$)  is assigned an identically distributed random variables $\eta(i,h)$ with normal Gaussian
distribution with zero mean and unit variance. The random disorder $f_p(x_i,u_i^n)$ is obtained by the linear interpolation of the random
force between two random variables $\eta(i,h)$ and $\eta(i,h+1)$ where $h = [u_i^n/\epsilon]$.

The numerical investigation of the scaling characteristics and critical exponents of the quenched generalized elastic model for different values of the fractional order $z$ and the non-local hydrodynamic interaction power $\alpha$ is presented in detail in the next section.

\begin{table*}[htp]
\caption{Measured exponents from numerical simulation of qGEM model with local hydrodynamic interaction $\alpha=1$ for different values of $z$.}
\label{tab1}
\begin{tabular}{|c|c|c|c|c|c|}
\hline
$z$              & $\theta$ & $\beta$ & $\zeta_g$ & $\zeta_l$ & $\sigma$ \\
\hline
$1.5$ & $0.512\pm 0.005$ & $0.877 \pm 0.002$ & $1.244\pm 0.003$ & $0.915\pm 0.005$ & $0.124\pm 0.003$  \\
$2.0$            & $0.445\pm 0.003$ & $0.875 \pm 0.003$ & $1.255\pm 0.005$ & $0.925\pm 0.004$ &  $0.125\pm 0.005$  \\
$2.5$           & $0.376\pm 0.004$ & $0.869 \pm 0.002$ & $1.263\pm 0.004$ & $0.935\pm 0.005$ &  $0.128\pm 0.002$  \\
$3.0$     & $0.315\pm 0.005$ & $0.862 \pm 0.002$ & $1.268\pm 0.003$ & $0.955\pm 0.003$ &  $0.134\pm 0.002$ \\
$3.5$     & $0.294\pm 0.004$ & $0.851 \pm 0.004$ & $1.302\pm 0.004$ & $0.990\pm 0.004$ &  $0.143\pm 0.003$  \\
$4.0$     & $0.285\pm 0.002$ & $0.835 \pm 0.003$ & $1.358\pm 0.004$ & $1.095\pm 0.005$ &  $0.155\pm 0.002$  \\
\hline

\end{tabular}
\end{table*}

\begin{table*}[htp]
\caption{Measured exponents from numerical simulation of qGEM model with $z=4$ and different values of non-local hydrodynamic interaction parameter $\alpha$.}
\label{tab2}
\begin{tabular}{|c|c|c|c|c|c|}
\hline
$\alpha$    & $\theta$ & $\beta$ & $\zeta_g$ & $\zeta_l$ & $\sigma$ \\
\hline
$1.0$ & $0.285\pm 0.002$ & $0.835 \pm 0.003$ & $1.358\pm 0.004$ & $1.095\pm 0.005$ &  $0.155\pm 0.002$  \\
$0.8$            & $0.297\pm 0.004$ & $0.704 \pm 0.002$ & $1.220\pm 0.003$ & $1.010\pm 0.005$ &  $0.334\pm 0.001$  \\
$0.6$           & $0.311\pm 0.005$ & $0.621 \pm 0.004$ & $1.107\pm 0.004$ & $0.985\pm 0.004$ &  $0.427\pm 0.003$  \\
$0.4$     & $0.331\pm 0.006$ & $0.556 \pm 0.004$ & $1.020\pm 0.004$ & $0.980\pm 0.003$ &  $0.466\pm 0.003$ \\
$0.2$     & $0.361\pm 0.006$ & $0.503 \pm 0.005$ & $0.954\pm 0.005$ & $0.970\pm 0.005$ &  $0.486\pm 0.002$  \\

\hline

\end{tabular}
\end{table*}

\section{Numerical results}\label{Numerical results sec}
To determine the time evolution of the interface specified by $u(x,t)$
and to obtain the critical properties of the qGEM, 
the simulation is started with initial condition $u(x,0) = 0$ , and boundary condition $u(x,t) = u(x+L,t)$.
We simulated this model on a lattice of size $L \in \left\lbrace 64, 128,256,512,1024,2048 \right\rbrace$. In addition, we carefully
choose the  time increment $\Delta t$ small enough to
ensure the stability of the numerical algorithm.

In order to determine the criticality of the qGEM (\ref{qGEM}) and (\ref{discretqGEM}) for various parameter
values of the fractional order $z$ and the hydrodynamic interaction parameter $\alpha$, we first compute the average velocity
$v(t,F) = \frac{d}{dt} \langle \int  u(x,t)dx \rangle$ as a
function of time for various values of the external homogeneous force $F$.

Surprisingly,
our simulations indicate that, the qGEM model in the limit $t\rightarrow \infty$ exhibits three quite different behaviours depending on the values of $z$ and $\alpha$. 
When hydrodynamic interactions is strongly long-range $\alpha \ll 1$ and the fractional power $z\ll 4$,  there exists
no phase transition between a pinned phase and a moving phase. In this regime $\lim _{t\rightarrow\infty} v(t,F)=0$ for an arbitrary external driving force $F$. Such a behaviour is shown on the top panel of Fig. (\ref{Figure:1}) for $\alpha =0.5$ and $z = 1.0$. 
In the opposite limit when the parameters $\alpha \leq 1 $ and $z \gg 1 $ the velocity of the interface remains zero (\textit{pinned} phase) up to a critical force $F_c$ and above $F_c$  
the velocity $v(t)$ decreases
as a power-law at the beginning and then becomes constant at all later time \textit{i.e.} $\lim _{t\rightarrow\infty} \frac{d}{dt} v(t,F)=0$ (\textit{moving} phase). 
As indicated in the bottom panel Fig. (\ref{Figure:1}), $v(F)$ is a continuous function of $F$. Thus the transition,
looks similar to the continuous phase transition in the context of the critical phenomena. 
Another surprising features of the qGEM model is the anomalous pinning-depinning transition for some specific values of the parameters $\alpha$ and $z$ in the ($\alpha$,$z$) plane. In the anomalous regime, an elastic interface which exhibits non-trivial phase transition behavior is pinned when $F<F_c$. But for $F > F_c$ we observe a jump in the average velocity as a function of $F$ (see Fig. (\ref{Figure:1})), may lead to a first order
phase transition in which the order parameter of the system changes discontinualy from zero to a finite value. Note that above $F_c$ the average velocity varies with time $\lim _{t\rightarrow\infty} \frac{d}{dt} v(t,F)\neq 0$, which is noticeably
 different from a standard pinning-depinning phase transition appears in
 the elastic interface models. In Fig. (\ref{Figure:2}) one may see a so-called phase diagram calculated
for the generalized elastic model with quenched disorder.

We here focus on one aspect of the
problem namely the scaling
behavior with characteristic critical exponents of the qGEM close to the depinning critical point. At the depinning threshold $F_c$, the depinned interface shows scaling behaviors in the global interface width $W\sim t^\beta$, in the early time region and the growth velocity of the average height $v(t)=d\bar{u}/dt\sim t^{-\sigma}$. Since $W\sim \bar{u}$, which results in a relation $v(t)\sim t^{\beta-1}$. Therefore, the exponents $\beta$ and $\sigma$ are not independent and the relation $\beta + \sigma=1$ occurs. In tables \ref{tab1} and \ref{tab2}, summarize our numerical findings for exponents $\beta$ and $\sigma$ for different values of control parameters $z$ and $\alpha$. Interestingly, the results are in a good agreement by the prediction $\beta + \sigma=1$.  

When the time exceeds the characteristic
time $t^{\times}_W \sim L^\nu$, the global interface width $W(L,t)$ reaches a saturation value $W_s(L)$. To determine the global roughness exponent $\zeta_g$ for qGEM model we use the scaling relation $W_s(L)\sim L^{\zeta_g}$. We obtain $\zeta_g$ from the double-log plot of the saturated surface width as a function of the
system size. In tables \ref{tab1} and \ref{tab2} we have shown the results for various
values of $z$ and $\alpha$. To evaluate the local roughness exponent $\zeta_l$, we calculated tow-point correlation function $C(r)$ (see Eq. \ref{coorelation function}). The log-log diagram of $C(r)$ versus $r$ gives the slope $\zeta_l$. Our computations are reported in tables \ref{tab1} and \ref{tab2}. It seems that the local roughness exponent does not change with
respect to the control parameters $\alpha$ and $z$ and it is nearly constant equal to unity.  Finally, to further investigate the scaling behavior of the qGEM model, we compute mean velocity $v(F)$ to determine of the scaling exponent $\theta$ numerically. As mentioned, in the steady state there is a scaling relation $v\sim (F-F_c)^\theta$. We measured the exponent $\theta$ using this scaling relation which values are reported in tables \ref{tab1} and \ref{tab2}.

\section{Conclusions}\label{Conclusion sec}
In this paper we have studied the the depinning transition of the elastic interface with non-local hydrodynamic interactions. As we mentioned earlier this model is called generalized elastic model in the presence of quenched disorder. We numerically studied different aspects of this model for different values of the fractional order $z$ and the non-local hydrodynamic interaction power $\alpha$.
We found that the behaviour of order parameter $v(F)$ as function of the external force $F$  highly depends on the values of $z$ and $\alpha$. There are three distinct phases in the phase space $(z-\alpha)$. In the small values of $z$ and $\alpha$ the order parameter vanishes and in the thermodynamic limit the steady-state order parameter approaches zero.
In opposite limit, where $\alpha \sim 1$ and $z>>1$, the model exhibits second-order phase transition and the order parameter $v(F)$ continuously changes from zero to none-zero values. And finally there is an additional phase with the order parameter changes discontinuously changes from zero to non-zero values, which is characterized by a first-order phase transition. We have analysed in detail the steady state of the model
in the second-order phase transition regime. Our model displays naturally scaling features near critical point $F_c$. We measured different scaling exponents as functions of $z$ and $\alpha$. Our results are in a good agreements with the well-known models.

\end{document}